\documentclass[sigconf]{acmart}

\pdfoutput=1

\usepackage[linesnumbered,ruled,vlined]{algorithm2e}

\newtheorem{theorem}{Theorem}

\newtheorem{assumption}{Assumption}

\allowdisplaybreaks

\acmConference[MobiHoc '25]{the 26th International Symposium on Theory, Algorithmic Foundations, and Protocol Design for Mobile Networks and Mobile Computing}{October 27--30, 2025}{Houston, USA}

\begin{document}

\title{Consensus-based Decentralized Multi-agent Reinforcement Learning for Random Access Network Optimization}

\author{Myeung Suk Oh}
\email{oh.746@osu.edu}
\affiliation{
  \institution{The Ohio State University}
  \city{Columbus}
  \state{Ohio}
  \country{USA}
}

\author{Zhiyao Zhang}
\email{zhang.15178@osu.edu}
\affiliation{
  \institution{The Ohio State University}
  \city{Columbus}
  \state{Ohio}
  \country{USA}
}

\author{FNU Hairi}
\email{hairif@uww.edu}
\affiliation{
  \institution{University of Wisconsin-Whitewater}
  \city{Whitewater}
  \state{Wisconsin}
  \country{USA}
}

\author{Alvaro Velasquez}
\email{alvaro.velasquez@colorado.edu}
\affiliation{
  \institution{University of Colorado Boulder}
  \city{Boulder}
  \state{Colorado}
  \country{USA}
}

\author{Jia Liu}
\email{liu@ece.osu.edu}
\affiliation{
  \institution{The Ohio State University}
  \city{Columbus}
  \state{Ohio}
  \country{USA}
}

\begin{abstract}
    With wireless devices increasingly forming a unified smart network for seamless, user-friendly operations, random access (RA) medium access control (MAC) design is considered a key solution for handling unpredictable data traffic from multiple terminals. However, it remains challenging to design an effective RA-based MAC protocol to minimize collisions and ensure transmission fairness across the devices. While existing multi-agent reinforcement learning (MARL) approaches with centralized training and decentralized execution (CTDE) have been proposed to optimize RA performance, their reliance on centralized training and the significant overhead required for information collection can make real-world applications unrealistic. In this work, we adopt a fully decentralized MARL architecture, where policy learning does not rely on centralized tasks but leverages consensus-based information exchanges across devices. We design our MARL algorithm over an actor-critic (AC) network and propose exchanging only local rewards to minimize communication overhead. Furthermore, we provide a theoretical proof of global convergence for our approach. Numerical experiments show that our proposed MARL algorithm can significantly improve RA network performance compared to other baselines. 
\end{abstract}


\maketitle

\section{Introduction}\label{sec:intro}

\begin{figure}[!t]
    \centering
    \includegraphics[width=0.95\linewidth]{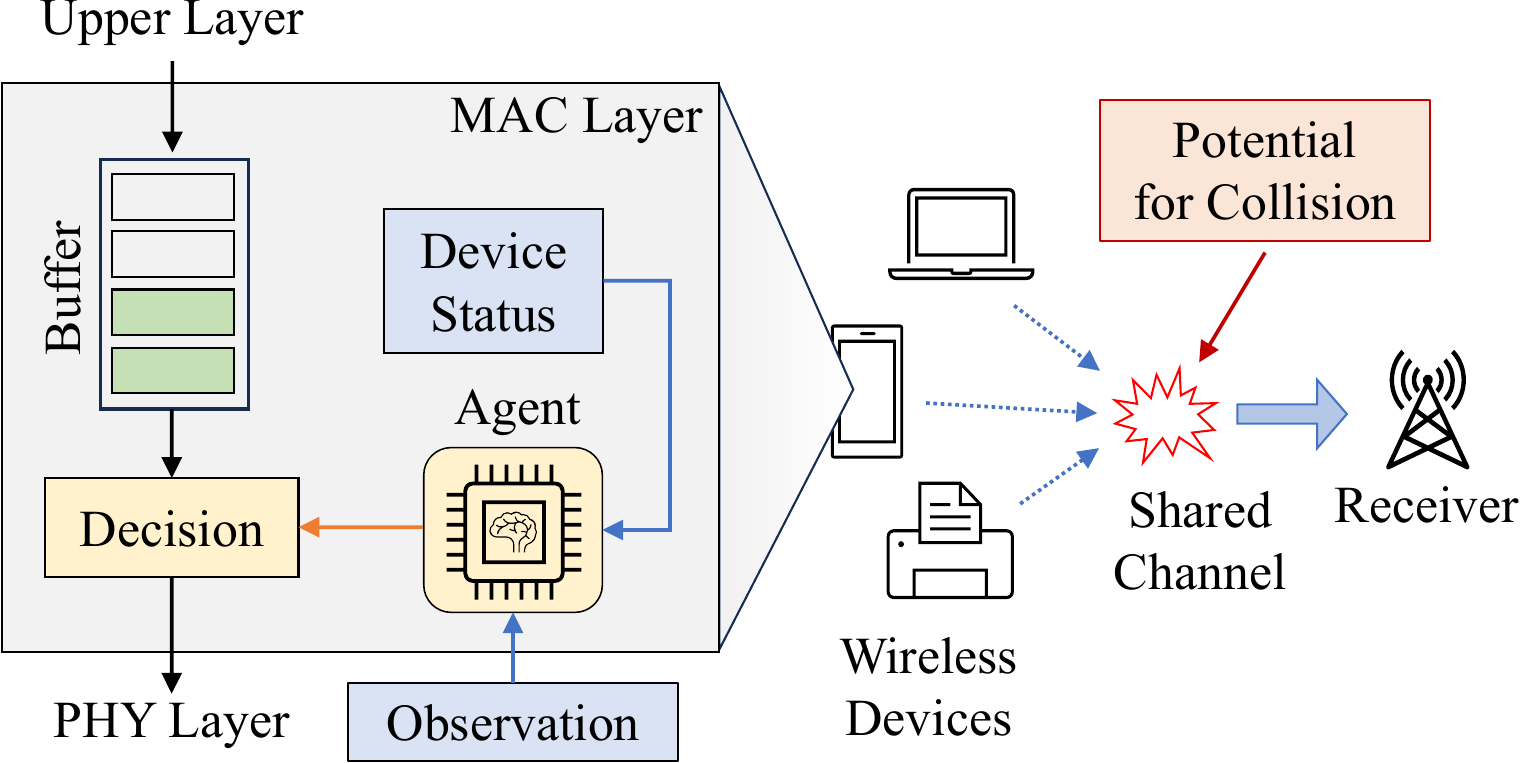}
    \caption{A visual illustration of random access (RA) network, where multiple wireless devices contend for access to a shared communication channel. Each device employs its own decision-making mechanism to determine when to transmit a packet stored in its buffer.}
    \Description{A visual illustration of random access (RA) network, where multiple wireless devices contend for access to a shared communication channel. Each device employs its own decision-making mechanism to determine when to transmit a packet stored in its buffer.}
    \label{fig:RA_structure}
    \vspace{-0.2in}
\end{figure}

\textbf{1) Background and Motivations:} Originating from the ALOHA protocol in the 1970s~\cite{Abramson70,Binder75} and going through the subsequent evolutions of carrier sensing multiple access (CSMA) technologies~\cite{Kleinrock75}, random access (RA) medium access control (MAC) design has been woven into the current fabric of the Internet, becoming an indispensable component of generations of Ethernet and Wi-Fi network standards (e.g., CSMA/CA widely adopted in Wi-Fi and Long Term Evolution Licensed Assisted Access (LTE-LAA)~\cite{Chen16} standards).
The sustained popularity of RA-based MAC is primarily due to its simplicity and flexibility in channel utilization when different user devices interact.
Specifically, in an RA network (shown in Figure~\ref{fig:RA_structure}), multiple devices share the same communication channel for data transmission.
The devices do not rely on centralized control for data traffic management (e.g., transmission scheduling handled by a server) but make independent decisions on when to transmit their data.
This \emph{decentralized} approach allows much simpler channel access management by eliminating the need for allocating a dedicated channel for each device.
RA is particularly effective in environments with a large number of devices that are connected with random data transmissions.
Moreover, through various emerging networking paradigms (e.g., Internet of Things (IoT)~\cite{Aouedi24}, machine-type communications~\cite{Mahmood21}, and smart grid communication infrastructure~\cite{Rodriguez23}), it is expected that RA-based MAC will continue to drive the development of future large-scale networks with seamless and user-friendly operations.

However, compared to its controlled access counterparts (e.g., TDMA, FDMA, and CDMA), the fundamental challenge in RA-based MAC lies in how to avoid and resolve collisions caused by the contentions of the network devices. 
In the context of Wi-Fi networks (shown in Figure~\ref{fig:RA_structure}), when two or more devices simultaneously attempt for data transmission, a collision occurs.
If not treated appropriately, excessive collisions could significantly decay the network performance by wasting scarce network resources and outweigh the benefits of using a RA-based MAC protocol.
Over the decades, a significant amount of effort has been dedicated to the optimization of RA-based MAC design.
Unfortunately, to date, the performance of RA network remains far from satisfactory.
In the literature, although there exists a large body of works on RA-based MAC optimization (e.g., throughput, delay, and fairness of RA networks), these works are either (i) too heuristic to provide any optimality guarantee, or (ii) too heavily rely on idealized analytical models that often fail to capture real-world complexities (see Section~\ref{sec:related} for more in-depth discussions).
Meanwhile, with the advent of the IoT era with increasingly ubiquitous network access, the importance of RA-based MAC design is poised to increase for years to come.
This widening gap between RA-based MAC optimization research and the rising demand for highly efficient RA networks motivates us to revisit this vital topic.

\smallskip
{\bf 2) The AI/ML Approach and Limitations:}
With the rapid advancement in fields of artificial intelligence (AI) and machine learning (ML) in recent years, AI/ML optimization strategies have been increasingly explored and applied in RA-based MAC design (e.g.,~\cite{Ahmad20,Kulin21,Cao21}).
The rationale for using ML/AI approaches is their potential to address the limitations of idealized analytical models in RA-based MAC research through data-driven insights, while capitalizing on the rapid progress in AI/ML.
Specifically, due to the decentralized nature of RA networks, an RA-based MAC optimization problem can be viewed as a distributed decision-making problem in the form of multi-agent reinforcement learning (MARL).
As a subfield of reinforcement learning (RL), MARL focuses on scenarios where multiple agents coexist in a shared environment.
These agents learn by making interactions and receiving feedback from the environment, and their ultimate goal is to find a joint policy that maximizes the overall reward.

Inspired by this architectural fit, several MARL-based approaches have been proposed to tackle the RA-based MAC optimization problem~\cite{Yu19,Yu22,Zhang20,Guo22,He24}.
These methods commonly adopt the centralized training and decentralized execution (CTDE) framework to securely achieve convergence in MARL.
In this framework, a deep neural network (DNN) for policy evaluation is trained centrally, where information from each device is aggregated to form a global representation.
Once trained, the network coordinates with locally distributed decision-making DNNs, helping each device to make independent decisions on its action.

While CTDE-based MARL has demonstrated its effectiveness in RA-based MAC optimization and has successfully enhanced network performance, its dependence on centralized tasks can render the framework rather impractical for many real-world applications that do not assume the presence of a central entity.
Even in networks that support such an entity, collecting the necessary information (e.g., local observations and actions) from all devices for centralized training can incur significant communication overhead as the network scales.
Moreover, some RA-based networking scenarios may prioritize data privacy and network security and thus do not prefer involving centralized tasks due to the risk of having increased vulnerability to potential attacks and reduced fault tolerance.

\smallskip
{\bf 3) Our Approach and Contributions:}
To overcome the challenges mentioned above, we take an {\bf decentralized} MARL approach to solve the RA-based MAC optimization problem.
Specifically, we consider a fully decentralized MARL architecture, where policy learning is executed without the aid of centralized tasks.
In particular, we propose to leverage the average consensus mechanism in which devices locally communicate with their neighbors for information exchange to facilitate global convergence in MARL.
We specifically design our fully decentralized MARL algorithm over actor-critic (AC) learning, which has been widely applied in RL tasks with various architectures (e.g., advantage actor-critic (A2C)~\cite{Sutton98}, deep deterministic policy gradient (DDPG)~\cite{Lillicrap15}, and proximal policy optimization (PPO)~\cite{Schulman17}).
We show that our consensus-based fully decentralized MARL framework achieves a provable finite-time convergence rate guarantee as the CTDE approaches, while avoiding the limitations of CTDE approaches in practice.
Our main contributions are summarized as follows.

\vspace{-.15em}
\begin{list}{\labelitemi}{\leftmargin=1.5em \itemindent=-0.0em \itemsep=-.0em}
    \item We formulate a new fully decentralized MARL framework for RA-based MAC optimization problem, where we carefully design our reward function with simple device parameters such that maximizing the reward naturally improves the total network throughput while ensuring fairness across all agents. 
    
    \item Unlike existing works that adopt a CTDE approach, our consensus-based fully decentralized MARL approach does not require centralized procedures but instead relies only on local information exchanges between neighboring devices to achieve global convergence.
    The proposed algorithm is thus applicable in RA-based MAC scenarios where (i) a centralized controller is not available; and (ii) scalability, privacy, and security become critical aspects.
    
    \item We present a theoretical analysis demonstrating that our fully decentralized AC algorithm with local reward sharing can converge to a fixed point. Our analysis provides finite-time convergence rates for both the actor and critic. A key distinction from existing analyses is that our analysis reflects on consensus solely applied to the local rewards.
    
    \item We conduct extensive numerical experiments to evaluate the performance of our consensus-based fully decentralized MARL algorithm. Through comparisons with baseline methods, we show that our algorithm significantly improves RA-based network performance while ensuring fairness across devices.
\end{list}

The remainder of this paper is organized as follows.
Section~\ref{sec:related} reviews related works on RA-based MAC optimization. 
In Section~\ref{sec:system}, we present the system model for our RA network. Section~\ref{sec:algorithm} provides implementation details of our consensus-based decentralized MARL algorithm. 
We conduct numerical experiments in Section~\ref{sec:experiment}, and Section~\ref{sec:conclusion} concludes the paper.

\section{Related Work}\label{sec:related}

In this section, we provide a high-level overview of (i) traditional RA-based MAC optimization and (ii) state-of-the-art AI/ML-based approaches for RA network optimization.

{\bf 1) Traditional RA-based MAC Optimization:}
RA techniques are generally categorized into two types: sensing-free and sensing-based.
In sensing-free RA, devices do not monitor the channel before transmitting data.
Protocols like ALOHA~\cite{Abramson70} and its variants (e.g., slotted ALOHA) belong to this category.
Since sensing-free RA allows a device to initiate data transmission even when the channel is already in use, the chance of collision remains high.
Nonetheless, sensing-free RA has been recognized as a promising strategy for satellite communications~\cite{Choudhury83,Yu23} and multi-hop mobile networks~\cite{Baccelli06}.
In contrast, sensing-based RA employs the listen-before-talk (LBT) mechanism, where each device monitors the channel for idleness before attempting data transmission.
Sensing-based RA thus significantly reduces the collision rate compared to sensing-free cases.
A widely used sensing-based RA protocol nowadays is CSMA/CA~\cite{Kleinrock75}.
In addition to the LBT mechanism, CSMA/CA incorporates a random backoff time that delays each transmission attempt for further preventing collisions.
A heuristic way of generating backoff times in practice is binary exponential backoff (BEB), in which the size of the contention window (i.e., the range from which the backoff time is randomly selected) doubles after each collision.

While CSMA/CA is effective at reducing collisions, its heuristic way of setting backoff times can lead to increased transmission delays and less efficient channel usage.
Moreover, it is unclear whether CSMA has any theoretical performance guarantee.
Rather surprisingly, the seminal work in \cite{Jiang10:CSMA_TIT} shows that CSMA can be {\em ``throughput-optimal''} if one can adjust the backoff time and contention window size appropriately based on the queueing backlog at each device.
Since then, early 2010s have witnessed an intensive line of research on ``queue-length-based adaptive CSMA'' for RA-based MAC optimization (see, e.g.,~\cite{Jiang10:CSMA_TIT,Jiang10:CSMA_TON,Jiang12:CSMA_TIT,Ghaderi13:CSMA_TON,Liew09:CSMA_Product_Form,Ni12:Q-CSMA} and their follow-ups).
However, many of these theoretical studies relied on somewhat idealistic analytical models that often fail to capture real-world complexities.
Moreover, many of these algorithms suffer from poor delay and fairness as the network size increases.

{\bf 2) AI/ML Approaches for RA-based MAC Optimization:}
To address the above challenges in adaptive CSMA/CA, several RL-based approaches have been proposed.
For example, authors in~\cite{Han20} proposed Q-learning-based contention window selection algorithms for each cooperative and non-cooperative setting to maximize the total throughput while satisfying the fairness constraints.
In~\cite{Lee24}, deep RL based on soft actor-critic (SAC) and long short-term memory (LSTM) models was utilized to dynamically adjust device waiting time and optimize network throughput.
These approaches have been proven effective in enhancing RA performance.
However, since CSMA/CA fundamentally depends on probabilistic transmissions of each device, performance improvement is still limited by its nature of stochastic operation.

Another approach in RA network optimization is to develop a deterministic transmission policy for each participating device, for which several MARL-based strategies have been proposed.
In~\cite{Yu19}, a deep Q-network was adopted to make transmission decisions for each RA device with an aim to maximize the generalized $\alpha$-fairness objective.
This approach was later extended to account for an imperfect wireless channel in which feedback signals for information collection can be corrupted~\cite{Yu22}.
The work in~\cite{Zhang20} employed a federated learning framework to implement distributed policy learning in RA networks, where each device is equipped with a DNN for decision-making.
Furthermore, QMIX and multi-agent PPO algorithms were explored in~\cite{Guo22} and~\cite{He24}, respectively, to implement MARL-based RA and improve network performance.
Although these methods have shown promising results in RA optimization, they utilize the CTDE framework, which requires the existence of a central entity capable of handling a large communication overhead for information collection.
This requirement may not be practical in many real-world RA scenarios, especially where security and data privacy are of great concern.
This motivates us to consider a fully decentralized architecture and develop a consensus-aided MARL algorithm that performs RA optimization in a more scalable and robust manner.

\section{System Model and Problem Formulation}\label{sec:system}

We consider a time-slotted RA scenario over a finite time horizon $T$, where each time slot is indexed by $t=1,2,\ldots,T$.
There are $N$ devices in the network contending for channel access to transmit data packets either to an intended receiver.
To model a fully decentralized network, we do not assume the presence of an entity that is responsible for tasks like global information collection and centralized training.
Instead, each device is aware of nearby devices and capable of exchanging information with them.
We use a graph $\mathcal{G} = (\mathcal{N},\mathcal{E})$ to represent the network, where $\mathcal{N}=\{1,2,\ldots,N\}$ is the set of devices and $\mathcal{E}$ denotes the edge set~\cite{Boyd06,Zhang18,Hairi22}.

Each device $i\in\mathcal{N}$ is equipped with an internal packet buffer of size $Q_\text{max}$.
When a packet arrives at the MAC layer, it is first queued in the buffer and transmitted on a first-come first-served basis.
Here, we use $0\leq q_i^{(t)} \leq Q_\text{max}$ to represent the number of packets in the buffer of device $i$ at time slot $t$.
Similar to CSMA/CA, our RA network operates on a LBT mechanism, where each device first checks on the channel status before transmitting any packets.
If there are packets in a device's buffer, i.e., $q_i^{(t)}>0$, the device enters the clear channel assessment (CCA) phase to check if the channel is idle.
If the channel remains idle for an amount of time that is sufficiently long, the device considers the channel to be clear and decides whether to transmit its packet.
If the device decides to \textit{wait}, it simply returns to the CCA phase without further action; otherwise, if the device chooses to \textit{transmit}, a single packet is transmitted over the channel.

\begin{figure}[!t]
    \centering
    \includegraphics[width=0.85\linewidth]{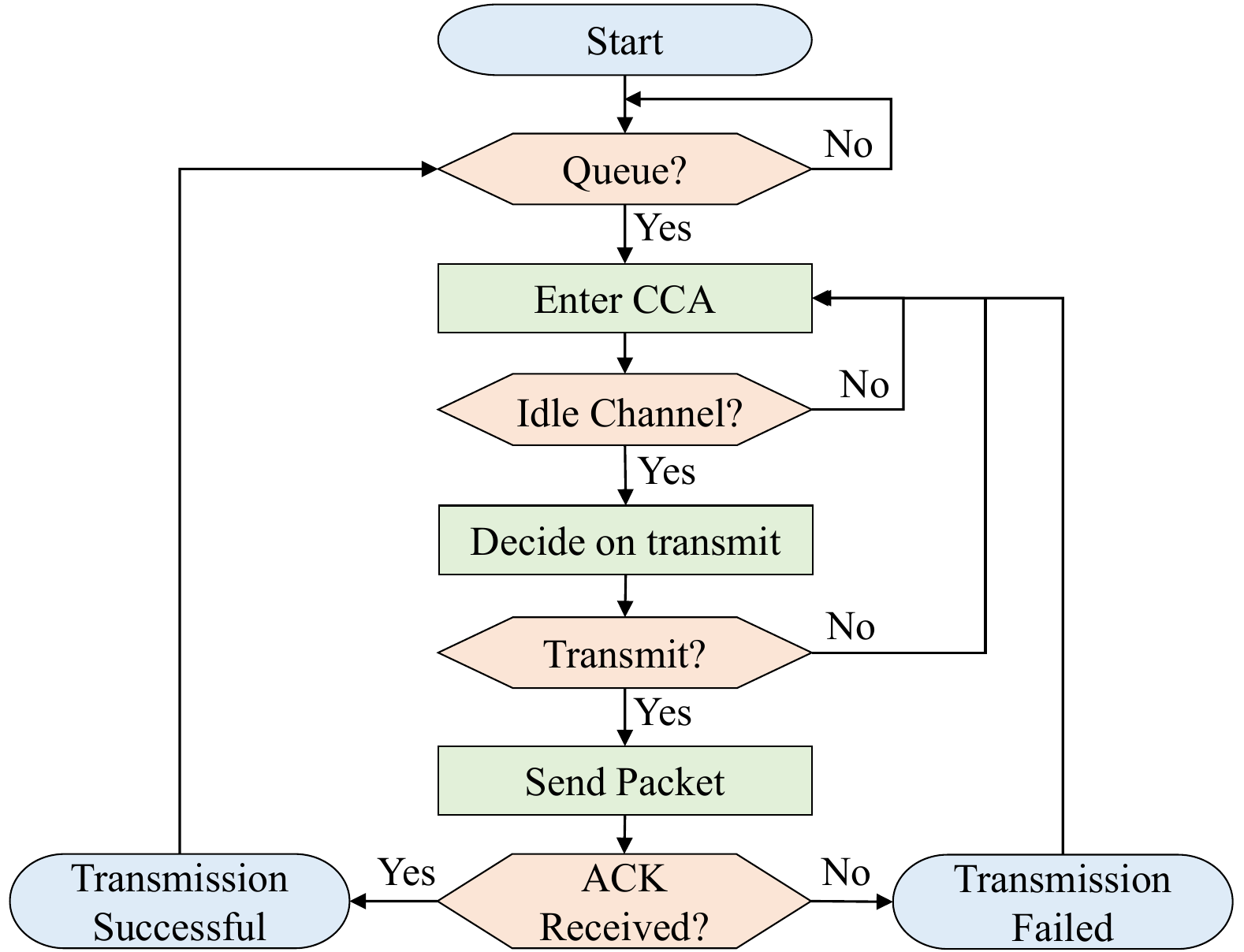}
    \caption{A flowchart showing the overall RA steps done by each device. The procedure follows the listen-before-talk (LBT) mechanism. The success of transmission is determined based on the reception of an ACK packet.}
    \Description{A flowchart showing the overall RA steps done by each device. The procedure follows the LBT mechanism. The success of transmission is determined based on the reception of an ACK packet.}
    \label{fig:flowchart}
\end{figure}

Upon a successful transmission (i.e., no collision occurs, and the packet is securely received), an acknowledgment (ACK) is sent by the receiver after some prefixed delay to confirm the successful transmission.
The device then returns to the CCA phase to prepare for transmitting the next packet.
In the event of collision, the intended receiver does not send an ACK, indicating a failed transmission.
Then, the device waits for the next opportunity to retransmit the packet.
In Figure~\ref{fig:flowchart}, we provide a flowchart summarizing the overall RA procedure.

To quantify the performance of our RA network, we define $v_i^{(t)}$ as the number of packets that have been successfully transmitted from device $i$ by time slot $t$.
We also define $l_i^{(t)}$ as the number of time slots elapsed since device $i$'s last successful transmission~\cite{Guo22}, which can be interpreted as the anticipated packet delay if device $i$ successfully transmits a packet at time slot $i$.
Our goal is to maximize the total network throughput, which is given by
\begin{equation}
    \max\frac{1}{T}\sum_{i=1}^{N} v_i^{(T)},
    \label{eq:throughput_1}
\end{equation}
while ensuring fairness across the devices.
Note that maximizing throughput can be simply achieved by assigning higher transmission priorities to particular devices such that collisions never occur.
However, such a policy may lead to significant imbalances in fairness.
The problem becomes non-trivial when both throughput and fairness must be considered simultaneously, i.e., achieving~\eqref{eq:throughput_1} while keeping $l_i^{(t)}$ at a similar level for all $i$.

\section{The Proposed Consensus-Based Fully Decentralized MARL Approach}\label{sec:algorithm}

In this section, we first present the fundamentals of MARL to provide background information.
Then, we provide our Markov decision process (MDP) formulation for RA-based MAC optimization.
Lastly, we will introduce our consensus-based decentralized MARL algorithm for RA-based MAC optimization.

\subsection{MARL: A Primer}\label{ssec:alg_MARL}

In MARL, each agent interacts with the environment through actions and observes the state and reward signals that represent the quality of the taken action.
Here, the actions performed by the agents are coupled and jointly impact the next state.
An MARL problem can be mathematically described by an MDP characterized by a 4-tuple:
    $\{\mathcal{S}, \{\mathcal{A}_i\}_{i\in\mathcal{N}}, P, \{R_i\}_{i\in\mathcal{N}}\}$,
where $\mathcal{S}$ is the global state space and $\mathcal{A}_i$ is the action set for agent $i$.
$P:\mathcal{S} \times \mathcal{A} \times \mathcal{S} \rightarrow [0,1]$ is the state transition probability, where $\mathcal{A} = \prod_{i\in\mathcal{N}} \mathcal{A}_i$ is the joint action set of all agents.
Lastly, $R_i:\mathcal{S} \times \mathcal{A} \rightarrow \mathbb{R}$ is the local reward function for agent $i$.

In a state $s^{(t)}\in\mathcal{S}$ at time slot $t$, each agent $i$ takes an action $a_i^{(t)}\in\mathcal{A}_i$ based on a policy $\pi_{\theta_i}$ parameterized by $\theta_i$.
Since the actions of agents are coupled, the state is transitioned to $s^{(t+1)}$ based on $P(s^{(t+1)}|s^{(t)},\boldsymbol{a}^{(t)})$ where $\boldsymbol{a}^{(t)}=[a_1^{(t)},a_2^{(t)},\ldots,a_N^{(t)}]$ is the joint action.
Similarly, the instantaneous local reward of agent $i$ for the action taken at time $t$ can be expressed as $r_i^{(t)}=R_i(s^{(t)},\boldsymbol{a}^{(t)})$.
Let $\theta = [\theta_1^{\top},\theta_2^{\top},\ldots,\theta_N^{\top}]^\top$be the joint weight vector of all $N$ policies, the objective of MARL is to find an optimal $\theta$ that maximizes the expected infinite-time discounted global reward, which can be written as:
\begin{equation}
    J(\theta) := \mathbb{E}_{\pi_{\theta}}\! \left[\sum_{t=0}^{\infty}\gamma^{t}\bar{r}^{(t)}\right],
    \label{eq:MARL_objective}
\end{equation}
where $\bar{r}^{(t)}=\frac{1}{N}\sum_{i=1}^{N}r_i^{(t)}$ is the global averaged reward and $\gamma\in[0,1]$ is the discount factor.
In MARL, a state value function is commonly used to evaluate a policy given by $\theta$, which can be defined as:
\begin{equation}
    V_{\pi_{\theta}} (s) = \mathbb{E}_{\pi_\theta}\left[\sum_{t=0}^{\infty}\gamma^t\bar{r}^{(t)}\Big|s^{(0)}=s\right].
    \label{eq:V_function}
\end{equation}
Since $V_{\pi_{\theta}}(s)$ is typically unknown, in the RL literature~\cite{Sutton98}, $V_{\pi_\theta}(s)$ is often estimated through a temporal differential (TD) learning process referred to as ``critic.''
Specifically, let $V_{w_i}(s)$ be the state value approximation function where $w_i$ is the parameter of agent $i$'s critic model.
Then, the TD learning is bootstrapped by using the Bellman optimality principle as follows:
\begin{equation}\label{eq:Q_function}
    V_{w_i} (s) = \mathbb{E}_{\pi_\theta}\Big[\bar{r} + \gamma V_{w_i}(s')\Big].
\end{equation}
In RL and MARL, the policy improvement (referred to as ``actor'') can be facilitated by updates based on the policy gradient:
\begin{equation}
    \nabla_{\theta_i}J(\theta) = \mathbb{E}_{s,\boldsymbol{a}} \Big[\nabla_{\theta_i}\log\pi_{\theta_i}(a_i|s)\cdot \text{Adv}_{\theta}(s,\boldsymbol{a})\Big],
    \label{eq:policy_gradient}
\end{equation}
where $\text{Adv}_{\theta}(s,\boldsymbol{a}) = \bar{r} + \gamma V_\theta (s') - V_\theta(s)$ is the advantage function.
Once the actor takes an action according to the current policy, the critic evaluates the policy based the acquired reward.
Upon evaluation, the policy is improved using~\eqref{eq:policy_gradient} so that the actor is able to take actions that lead to improved expected reward.

Compared to single-agent RL, the key differences in MARL are as follows~\cite{Feriani21}. 
%\zhiyao{ZY: Why do we mention technical challenges here in preliminaries?}
First, since multiple agents make independent decisions simultaneously, the environment is never seen as stationary to an action of each individual agent.
Second, due to the decentralized architecture, each agent may observe only a part of information available in the environment.
Therefore, careful design of MARL framework is essential to achieve performance comparable to that of centralized learning.

\subsection{The MARL Problem Formulation for RA-Based MAC Optimization}\label{ssec:alg_MDP}

With the MARL preliminaries, we are now in a position to formulate our RA-based MAC optimization as an MARL problem.
Toward this end, we first define the state in each time slot as:
\begin{equation}
    s^{(t)}=\left(\left\{\overline{q}_i^{(t)}\right\}_{i=1}^{N}, \left\{\overline{l}_i^{(t)}\right\}_{i=1}^{N}, c^{(t)}\right),
    \label{eq:MARL_state}
\end{equation}
where $\overline{q}_i^{(t)}=q_i^{(t)}/Q_\text{max}$ is the normalized packet queue of device $i$ at time slot $t$, $\overline{l}_i^{(t)}=\omega_0l_i^{(t)}$ is the normalized time delay since device $i$'s last successful transmission with $\omega_0$ being the scaling parameter, and $c^{(t)}\in\{0,1\}$ is the channel usage indicator.
We formulate our state such that the entire status of our RA network is accurately perceived.
In practice, each agent $i$ can only observe the part of state $s^{(t)}$, which we denote using $o_i^{(t)}\in s^{(t)}$ and define as follows:
\begin{equation}
    o_i^{(t)}=\left\{\overline{l}_i^{(t)}, \left\{\overline{l}_{j}^{(t)}\right\}_{j\in\mathcal{N}\setminus i}, c^{(t)}\right\}.
    \label{eq:MARL_obs}
\end{equation}
Note that, for each device $i$, the first component $\overline{l}_i^{(t)}$ is local information.
% \begin{equation}
%     o_i^{(t)}=\left\{\overline{q}_i^{(t)},\overline{l}_i^{(t)}, \left\{\overline{l}_{j}^{(t)}\right\}_{j\in\mathcal{N}\setminus i}, c^{(t)}\right\}.
%     \label{eq:MARL_obs}
% \end{equation}
% Note that, for each device $i$, the first two components $\overline{q}_i^{(t)}$ and $\overline{l}_i^{(t)}$ are local information.
As assumed in~\cite{Yu19,Guo22,He24}, we consider the time delays from other network devices, i.e., $\left\{\overline{l}_{j}^{(t)}\right\}_{j\in\mathcal{N}\setminus i}$ to be observable since they are traceable by listening to ACK packets broadcast by the receiver.
The last parameter $c^{(t)}$ is easily observable from monitoring the channel during the CCA phase.

For all devices, we use a discrete action space $\mathcal{A}_i=\{0, 1\}$, where $0$ represents the action of \textit{wait} and $1$ represents the action of \textit{transmit}.
In other words, we consider each device $i$ can take one of the two discrete actions at time $t$, i.e., $a_i^{(t)}\in\{0, 1\}$.

Similar to~\cite{Yu19,Guo22,He24}, we assume that each agent can store the $M$ latest observations and actions and use them as a set of observation-action history.
Let $\tilde{t}_{i,m}$ be the time when the $m$-th latest action was taken by device $i$.
Then, the observation-action history of length $M$ for device $i$ can be formed as
\begin{equation}
    \hspace{-0.5mm}\eta_{i,M:1}^{(t)} = \left\{o_i^{(\tilde{t}_{i,M})},a_i^{(\tilde{t}_{i,M})},o_i^{(\tilde{t}_{i,M-1})},a_i^{(\tilde{t}_{i,M-1})},\ldots,o_i^{(\tilde{t}_{i,1})},a_i^{(\tilde{t}_{i,1})}\right\}\hspace{-0.5mm}.\hspace{-0.5mm}
    \label{eq:MARL_hist}
\end{equation}
We aim to utilize~\eqref{eq:MARL_hist} as an additional information in feeding both actor and critic to make their learning process to reflect the dynamics of RA environment upon deciding and evaluating the action.

We define our instantaneous local reward for device $i$ at time slot $t$ to be
\begin{equation}
    r_i^{(t)} = -\left(\omega_1\overline{l}_i^{(t)}+\omega_2\overline{q}_i^{(t)}\right),
    \label{eq:MARL_reward}
\end{equation}
where $\omega_1$ and $\omega_2$ are scaling factors.
We strictly define our reward function to be local, i.e., $r_i^{(t)}$ is not a function of the information from other agents, to reflect the condition of fully decentralized MARL.
For the consensus step, we let $r_i^{(t)}$ to be exchanged across the devices through local communication links defined by $\mathcal{G}$.

As outlined in~\eqref{eq:MARL_objective}, the objective of our MARL is to maximize the long-term discounted global reward.
When we consider~\eqref{eq:MARL_reward} for the reward function in~\eqref{eq:MARL_objective}, we observe that our MARL is designed to focus on minimizing both the packet delays and the packet queues within the RA network.
Unlike the rewards defined in~\cite{Yu19,Guo22,He24}, which assume through CTDE framework that either (i) devices have access to the global reward or (ii) the global reward is directly computed by the central entity, we avoid incorporating action-dependent scores (e.g., assigning negative values upon collision) but instead use status-dependent scores to formulate our reward.
This is to prevent inefficient learning that may result from correlating locally exchanged rewards to each device's local action only in a simplistic manner.
We rather aim for our global reward $\bar{r}^{(t)}$ to focus on reflecting the condition of the RA network, which can directly be interpreted as the state value in our AC learning framework.

Moreover, we adopt our reward design in~\eqref{eq:MARL_reward} for the following reasons. 
Considering a finite time-horizon $T$, let $\mathcal{X}_i^{(T)}$ denote the set of time slots at which device $i$ successfully transmits a packet over the $T$ time slots.
Then, the network throughput as given in~\eqref{eq:throughput_1} can be rewritten as
\begin{equation}
    \frac{1}{T}\sum_{i=1}^{N}v_i^{(T)}=\sum_{i=1}^{N}\frac{|\mathcal{X}_i^{(T)}|}{\sum_{t\in\mathcal{X}_i^{(T)}} l_i^{(t)}}.
    \label{eq:throughput_2}
\end{equation}
Note that the denominator is fixed at $T$ regardless of how often successful transmissions occur.
Therefore, the only factor influencing throughput is the numerator, which can be maximized by increasing the number of successful transmissions.
Next, packet queues are included as part of the reward to ensure that our MARL reflects the need to prioritize devices with high transmission urgency and thus prevent packet loss due to buffer saturation.
Since the queue size is not included in the observation $o_i^{(t)}$, we allow the MARL to reflect each device's urgency in a stochastic manner for decision-making without directly correlating it with a deterministic action.

\subsection{The Proposed Consensus-Based Decentralized MARL Algorithm}\label{ssec:alg_implementation}

\begin{figure}[!t]
    \centering
    \includegraphics[width=0.95\linewidth]{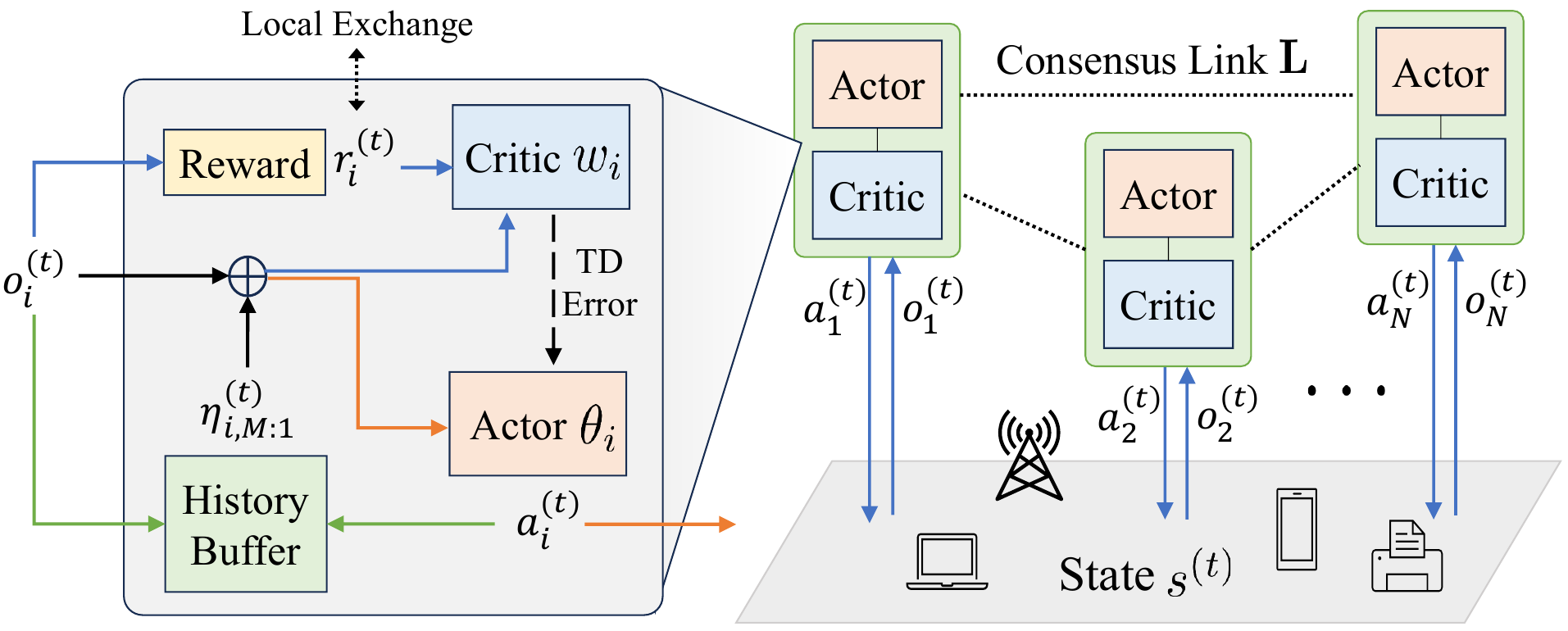}
    \caption{A visual representation of our fully decentralized MARL framework. Both the actor and critic are trained in a decentralized manner. There exist local communication links among devices that allow local reward sharing for consensus.}
    \Description{asdf}
    \label{fig:MARL}
    \vspace{-.2in}
\end{figure}

The overall architecture of our consensus-based fully decentralized MARL for RA network optimization is illustrated in Figure~\ref{fig:MARL}.
Each device $i\in\mathcal{N}$ updates its own AC models trained using local experiences.
To store and use the observation-action history at each learning step, each device maintains a history buffer to record the past observation-action pairs.
As described in Section~\ref{sec:system}, connected devices can exchange their local information with each other.
In our consensus-based decentralized MARL algorithmic design, we allow each node to share local rewards with its neighbors.

Our proposed consensus-based decentralized MARL for the RA-based MAC layer of an $N$-device network is summarized in Algorithm~\ref{alg:main}.
As discussed in Section~\ref{sec:system}, each device obeys the LBT mechanism (Figure~\ref{fig:flowchart}) and constantly monitors the channel.
Once the channel is assessed to be clear, the device takes an observation $o_i^{(t)}$ and makes a transmission decision $a_i^{(t)}$ using its local policy conditioned on the stored observation-action history and current observation \textbf{(Lines 6 - 11)}.
Depending on the transmission result (whether or not collision has occurred), each device updates its status.
The above step is repeated for the span of $T$ time slots.

Since most of time slots are occupied by the RA protocol steps such as CCA, packet transmission, and waiting upon ACK, the actual time slots related to MARL procedure are confined.
Hence, we consider our MDP to only progress over time slots where an action is taken by the devices.

Each time the devices acquire enough information to perform the MARL step (i.e., weight parameter update using gradient descent), the {\em consensus process}  \textbf{(Lines 19 - 22)} first starts as an initial step.
Each consensus step consists of $G$ rounds of communication, where weight-based averaging is performed in each round.
Then, each device updates its actor and critic parameters after computing the TD error \textbf{(Lines 24 - 27)}.
As a final step, both the current observation and action are stored in the history buffer.

\begin{algorithm}[!t]
	\caption{The Consensus-based Fully Decentralized MARL for the RA-based MAC Layer Optimization.}
    \label{alg:main}
    \textbf{Input:} agent set $\mathcal{N}$, neighbor sets $\{\mathcal{N}_i\}_{i\in\mathcal{N}}$, time horizon length $T$, history length $M$, consensus weight matrix $\mathbf{L}$, consensus iteration count $G$, actor rate $\alpha$, critic rate $\beta$ \\
    \textbf{Initialize:} actor weights $\theta_i$, critic weights $w_i$, transmit flag $\text{f}_i=\text{False}$, and ready-for-update status $u_i=\text{False}$, $\forall i\in\mathcal{N}$ \\
    \For{$t=1, 2 \dots, T$}{
        \For{$i \in \mathcal{N}$}{
            Update $q_i^{(t)}$, $l_i^{(t)}$, $c^{(t)}$, and $\text{f}_i$  \\
            \If{$q_i^{(t)}>0$ \& $c^{(t)}=0$}{
                Acquire observation $o_i^{(t)}$ and reward $r_i^{(t)}$\\
                Select $a_i^{(t)}\sim\pi_{\theta_i}(\:\cdot\:|\{\eta_{i,M:1}^{(t)},o_i^{(t)}\})$ \\
                \If{$a_i^{(t)}=1$}{
                    $\text{f}_i \leftarrow \text{True}$
                }
                $u_i \leftarrow \text{True}$ \\
            }
        }
        \For{$i \in \mathcal{N}$}{
            \If{$\text{f}_i = \text{True}$}{
                $c^{(t)} \leftarrow 1$ \\
                Transmit a packet
            }
            \If{ACK received}{
                $l_i^{(t)} \leftarrow 0$
            }
        }
        \If{$u_i=\text{True}, \forall i\in\mathcal{N}$}{
            $\tilde{r}_{i,0} \leftarrow r_i^{(t)}$ for all $i\in\mathcal{N}$ \\
            \For{$g=1, 2 \dots, G$}{
                $\tilde{r}_{i,g} \leftarrow \sum_{j\in\mathcal{N}_i} \ell_{ij}\tilde{r}_{j,g-1}$ for all $i\in\mathcal{N}$
            }
            $\tilde{r}_{i}^{(t)} \leftarrow \tilde{r}_{i,G}$ for all $i\in\mathcal{N}$ \\
            \For{$i \in \mathcal{N}$}{

                $\delta_i\hspace{-1mm} \leftarrow \tilde{r}_{i}^{(t)} + \gamma V_{w_i}\big(\{\eta_{i,M:1}^{(t)},o_i^{(t)}\}\big) - V_{w_i}\big(\{\eta_{i,M+1:2}^{(t)},o_i^{(\tilde{t}_{i,1})}\}\big)$ \\ 
                $w_i \leftarrow w_i - \beta \delta_i \cdot \nabla V_{w_i}\big(\{\eta_{i,M+1:2}^{(t)},o_i^{(\tilde{t}_{i,1})}\}\big)$ \\

                $\delta_i\hspace{-1mm} \leftarrow \tilde{r}_{i}^{(t)} + \gamma V_{w_i}\big(\{\eta_{i,M:1}^{(t)},o_i^{(t)}\}\big) - V_{w_i}\big(\{\eta_{i,M+1:2}^{(t)},o_i^{(\tilde{t}_{i,1})}\}\big)$ \\
                $\theta_i \leftarrow \theta_i + \alpha\delta_i \cdot \nabla\log\pi_{\theta_i}(a_i^{(\tilde{t}_{i,1})}\vert\{\eta_{i,M+1:2}^{(t)},o_i^{(\tilde{t}_{i,1})}\})$ \\

                Store $o_i^{(t)}$ and $a_i^{(t)}$ in the history buffer \\
                $u_i \leftarrow \text{False}$ \\
            }
        }
    }
    \textbf{Output:} $\theta_i$ for all $i\in\mathcal{N}$ \\
\end{algorithm}

A key {\bf novelty} of our algorithm is that we only exchange the local reward $r_i^{(t)}$ during the consensus step.
Note that this differs from many existing decentralized MARL algorithms (e.g.,~\cite{Zhang18,Chen22,Hairi22,Hairi24}), where the entire weights of critic (i.e., $w_i$) are exchanged during each consensus step to ensure convergence.
Given that DNNs are often employed to estimate the value function, the number of parameters that need to be exchanged between agents is huge.
In RA networks, this can be an unrealistic requirement since resource allocated for each established link can often be much limited.
By contrast, in our algorithm, each agent only needs to exchange its reward value in each time slot, which is only a {\em scalar}.
Thus, our approach significantly reduces the amount of information exchanged across the network and renders our approach more practical.
In Section~\ref{ssec:alg_analysis}, we will rigorously show the convergence performance of Algorithm~\ref{alg:main}.
We also note that our proposed algorithm could also of independent interest in the MARL literature.

\begin{figure}[!t]
    \centering
    \includegraphics[width=0.95\linewidth]{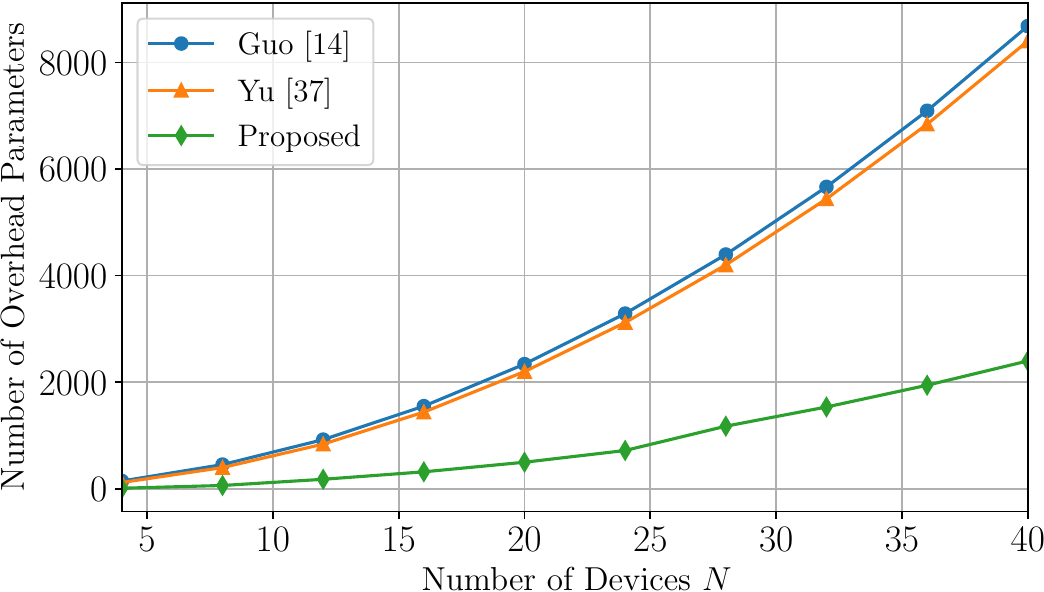}
    \caption{A plot comparing the number of overhead parameters required to perform MARL. CTDE approaches require more parameters to be exchanged than our consensus-based decentralized method.}
    \Description{asdf}
    \label{fig:overhead}
\end{figure}

To further demonstrate the above advantage of our algorithm, we compare our algorithm with one of CTDE in the amount of information exchanged for each consensus step.
In CTDE, a central entity must collect information from all participating devices to train the centralized critic, which usually takes a set of history, currently taken observation-action pair, and reward from every device to compute the gradient and update its weights.
Let us define $D_o$, $D_a$, and $D_r$ to be the dimensions of observation, action, and reward, respectively, for each given algorithm.
Then, for centralized critic, the total number of parameters that must be collected for each learning step is given by $N[M(D_o+D_a)+D_r]$.
On the other hand, our consensus-based approach requires each agent to exchange information with its neighboring devices, for which a total number of $NKGD_r$ parameters need to be exchanged, where $K$ is the average number of links per device.
We can specifically set $G=\big\lceil \frac{0.5\log\epsilon^{-1}}{\log\lambda_2^{-1}(\mathbf{L})} \big\rceil$ to achieve the normalized root mean squared error (RMSE) of $\epsilon$ for a given consensus weight matrix $\mathbf{L}$~\cite{Boyd06}.
We use two of the existing CTDE works~\cite{Guo22,Yu22}, where we provide the values of $D_o$, $D_a$, $D_r$ in Table~\ref{tab:overhead}, to compare the total number of required overhead to support a given number of devices.
According to the result shown in Figure~\ref{fig:overhead}, where we consider $\epsilon = 0.005$ and $K=N/4$, we can see that the fully decentralized case with local information exchange requires much {\em less} overhead to conduct MARL.

\begin{table}[!t]
    \centering
    \setlength\tabcolsep{4pt}
    \caption{Dimension of MDP Parameters for Different CTDE-based RA Algorithms.}
    \begin{tabular}{c|c|c|c}
        \hline
        Algorithm & $D_o$ & $D_a$ & $Dr$ \\
        \hline
        Guo~\cite{Guo22} & $N+2$ & $1$ & $2$ \\
        Yu~\cite{Yu22} & $1$ & $1$ & $NM$ \\
        \hline
    \end{tabular}
    \label{tab:overhead}
\end{table}

\subsection{Theoretical Performance Analysis}\label{ssec:alg_analysis}

Before providing our theoretical results regarding Algorithm~\ref{alg:main}, we provide some necessary assumptions in the following.

\begin{assumption}[Bounded Reward] \label{ass:r-max}
    There exists a positive constant $r_{\text{max}}$ such that $r_i^{(i)} \in [-r_{\text{max}}, r_{\text{max}}]$ for any $t\ge 0, i\in\mathcal{N}$.
\end{assumption}

\begin{assumption}[Mixing Time]\label{ass:mixing}
     There exist a stationary distribution $\zeta$ for $(s,a)$, and positive constants $\kappa$ and $\rho\in(0,1)$, such that $\sup_{s\in\mathcal{S}}\|P(s^{(t)},a^{(t)}|s^{(0)}=s) - \zeta(\theta)\|_{TV}\le \kappa\rho^t, \forall t\ge 0$.
\end{assumption}

\begin{assumption}[Lipschitz Continuity]\label{ass:smoothness}
    $J(\theta)$ is $L_J$-Lipschitz continuous w.r.t. $\theta$, i.e., there exists a positive constant $L_J$ such that, for any $\theta$ and $\theta'$, we have $|J(\theta)-J(\theta')| \le L_J\|\theta-\theta'\|_2$.
\end{assumption}

\begin{assumption}[Consensus Matrix]
    The consensus weight matrix $\mathbf{L}$ is doubly stochastic. Additionally, for all $i,j\in \mathcal{N}$, there exists a positive constant $\nu > 0$ such that (i) $\ell_{ii} \geq \nu$ and (ii) $\ell_{ij} \geq \nu$ whenever devices $i$ and $j$ are connected.
    \label{ass:consensus}
\end{assumption}

Now, we are ready to state our main theoretical results.

\begin{theorem}[Finite-Time Critic Convergence Rate]
    Consider the iterative updates of $w_i^{(t)},\forall i\in{\mathcal{N}}$, in Algorithm~\ref{alg:main}. For any given policy $\pi_\theta$ and $i\in\mathcal{N}$, it holds that
    \begin{align}
        \mathbb{E}\Big[\|w_i^{(t)} - w_{i}^{(t)*}\|^{2}\Big]
        \leq 2&\left(\frac{c_1}{c_3}\right)^2e^{-2c_2G + 2c_4t} \nonumber \\
        &+ 2 c_5(1-c_6\beta)^{t-\tau(\beta)}+ 2c_7\tau(\beta)\beta, \label{eq:critic}
    \end{align}
    where $c_1=2\beta N r_\text{max}(1+\nu^{-(N-1)})$, $c_2 =\ln(1-\nu^{N-1})^{-1}$, $c_3=(1+\gamma)\beta$, $c_4=\ln(1+(1+\gamma)\beta)$, $\{c_5, c_6, c_7\}$ are constants independent of the step size $\beta$, and $\tau(\beta) = \mathcal{O}(\log (\beta^{-1}))$ is the mixing time.
    \label{thrm:critic}
\end{theorem}
\textit{Proof Sketch}. To bound the difference between $w_i^{(t)}$ and $w_i^{(t)*}$, we decompose it as $w_i^{(t)}-w_i^{(t)*}=(w_i^{(t)}-\overline{w}_i^{(t)})+(\overline{w}_i^{(t)}- w_i^{(t)*})$, where $\overline{w}^{(t)}_i \triangleq \frac{1}{N}\sum_{i=1}^N w^{(t)}_i$.
For the first term, we derive its iterative expression and show that it depends on the initial consensus error in both the local reward and critic parameters.
We then prove that the error magnitude remains bounded under appropriately chosen learning parameters.
The analysis of the second term follows the approach provided in~\cite{Srikant19}, which handles the case of single-agent AC.
Due to space limitation, we relegate the full proof to our online technical report \cite{proof}
\qed 

We remark that the above result establishes convergence to the TD fixed points for all agents even when only rewards are shared rather than critic parameters as in related works~\cite{Zhang18,Chen22,Hairi22}.
From the first term in \eqref{eq:critic}, we see that the communication rounds $G$ for reward sharing must be sufficiently large.
For the sample complexity, we ignore logarithmic factors for simplicity by using the canonical $\tilde{O}(\cdot)$ notation.
To make the right-hand side of \eqref{eq:critic} as $\mathcal{O}(\epsilon)$ for any target threshold $\epsilon > 0$, we require $\beta = \tilde{\mathcal{O}}(\epsilon)$, $t = \tilde{\mathcal{O}}(\epsilon^{-1})$ and $G = \tilde{\mathcal{O}}(\epsilon^{-1})$.
As a result, the sample complexity for the critic is $t = \tilde{\mathcal{O}}(\epsilon^{-1})$ while the overall sample complexity is $tG = \tilde{\mathcal{O}}(\epsilon^{-2})$.
This sample complexity is on the same order of that in \cite{Xu20,Chen22,Hairi22}.
Regarding communication complexity, our result of $\mathcal{O}(\epsilon^{-2})$ may look worse than the aforementioned works.
For example, \cite{Chen22} claims to have $O(\log(\epsilon^{-1}))$. 
However, note that these prior works share a $d$-dimensional critic parameter whereas we only exchange scalar rewards.
Hence, the scalar communication complexity of \cite{Chen22} requires $O(d\log(\epsilon^{-1}))$ versus our $O(\epsilon^{-2})$. If $d = \Omega(\epsilon^{-2})$, our method actually performs more efficiently.
For instance, in MDPs with a very large state space $|\mathcal{S}|$, it often requires $d$ to be high-dimensional. 
If we choose $d = 10^3$ then with $\epsilon = 0.1$, our proposed algorithm provides an order-wise improvement in communication efficiency.

\begin{theorem}[Convergence Rate of Decentralized MARL with Local Reward Consensus]
    Consider the AC algorithm in Algorithm~\ref{alg:main}. With step-size set as $\alpha=\frac{1}{4L_J}$, it holds that
    \begin{align}
        \mathbb{E}\Big[\|\nabla_\theta J(\theta&^{(\hat{T})})\|^2\Big] \leq \frac{16L_J r_\text{max}}{T(1-\gamma)} \nonumber \\ 
        & + 18N(1+\gamma)^2 \frac{\sum_{t=1}^T\|w_i^{(t)}-w_i^{(t)*}\|^2}{T} \nonumber \\
        & + 72N^3r_\text{max}^2\left((1+\nu^{-(N-1)})(1-\nu^{N-1})^{G}\right)^2 \nonumber \\
        & + 18(1+\gamma)^2\xi^\text{critic}_\text{approx} + 72N(r_\text{max} + (1+\gamma)R_w)^2  \label{eq: actor},
    \end{align}
    where $\hat{T}$ is sampled uniformly from $\{1,\cdots,T\}$ and $R_w$ is a constant that is independent of $T$.
    \label{thrm:actor}
\end{theorem}
\textit{Proof Sketch}. Using the Lipschitz property, we first apply the descent lemma to express the gradient step in inequality form.
After rearranging the terms, we derive the bound on $\|\nabla_\theta J(\theta^{(t)})\|^2$ and decompose it into multiple error components.
We then upper bound each component and show that its magnitude can be efficiently controlled to converge over the given time steps $T$.
For the critic approximation error, we apply Theorem~\ref{thrm:critic} to show its convergence.
For the error that is due to imperfect reward consensus, we show its convergence using the doubly stochastic property of consensus weight matrix.
Due to space limitation, the full proof of Theorem~\ref{thrm:actor} is relegated to~\cite{proof}.
\qed

Based on Theorem~\ref{thrm:actor}, we ensure that the output policy of our Algorithm~\ref{alg:main} converges to the neighborhood of some stationary point at a rate of $\mathcal{O}(1/T)$.

\section{Numerical Experiments}\label{sec:experiment}

\subsection{Experimental Settings}\label{ssec:exp_setting}

We conduct numerical experiments to evaluate our proposed MARL algorithm for optimizing the IEEE 802.11 CSMA/CA MAC layer.
We consider $N=4$ devices participating in RA over $T=600$ time slots.
As described in Section~\ref{sec:system}, all devices operate under the LBT mechanism and follow the RA steps shown in Figure~\ref{fig:flowchart}.
Following the IEEE 802.11 protocol, we set the SIFS and DIFS durations to $2$ and $4$ time slots, respectively, where each slot lasts 9 $us$.
We assume each data packet is $1500$ bytes and requires $10$ time slots for transmission.
We consider ACK signals to take $4$ time slots and set the packet buffer size $Q_\text{max}=10$.
Packets arrivals are assumed to follow a Poisson point process of rate $\frac{1}{30}$ for each device.

The network graph $\mathcal{G}$ is generated using the Watts-Strogatz model~\cite{Watts98}, where each device connects to one neighboring device with zero rewiring probability.
In generating the consensus weight matrix $\mathbf{L}$, we assign equal weights to each device's established links, i.e., for each device $i\in\mathcal{N}$, $\ell_{ij}=\frac{1}{\vert\mathcal{N}_i\vert+1},\forall j\in\mathcal{N}_i$.
We set each consensus step consists of $G=3$ communication rounds.

For our consensus-based MARL algorithm, both actor and critic are implemented as multi-layer perceptrons (MLPs) with the width of $128$ and depth of $5$.
While ReLU activation is applied in each layer of our actor network, no activation function is used for our critic network to satisfy the linear value function approximation, i.e., $V_{w_i}(\cdot)=\phi^\top(\cdot)w_i,\forall i\in\mathcal{N}$, as required by our theoretical analysis.
The observation-action history length is set to $M=4$ to capture a moderate amount of past information.
We set $\omega_0=\frac{1}{60}$, $\omega_1=1$, and $\omega_2=1$ for MDP parameter scaling.
We use stochastic gradient descent (SGD) for both actor and critic weight updates with learning rates $\alpha=0.006$ and $\beta=0.003$, respectively.
We train our network over $1200$ episodes and take the average over $20$ independent runs.

For performance comparison, we consider the following RA baselines:
\begin{list}{\labelitemi}{\leftmargin=1.5em \itemindent=-0.0em \itemsep=-.0em}    
    \item \textbf{RA with a fixed transmission probability (RA-P)}: As considered in~\cite{Yu19,He24}, we consider a legacy RA protocol where each device transmits its packet based on a fixed transmission probability.
    According to the analysis in~\cite{Dai12}, the maximum throughput is achieved when the probability is set to $\frac{1}{N}$, which we set for our experiments.
    
    \item \textbf{RA with a fixed contention window (RA-FCW)}: Each device uses a backoff time randomly generated from a contention window of fixed size $W_\text{cw}$.
    We assume that the optimal value of $W_\text{cw}$ is previously found via experiments, i.e., we set $W_\text{cw}=16$ for our RA scenario of $M=4$ devices.
    
    \item \textbf{RA with an adaptive contention window (RA-ACW)}: Each device employs the BEB mechanism, where the size of the contention window doubles after each collision.
    For each device, we set the initial size of contention window as $W_\text{cw}=1$.
    
    \item \textbf{RA with an adaptive contention window (RA-CTDE)}: We consider an MARL algorithm with CTDE architecture.
    For the AC framework, CTDE is realized through a central critic, which collects information on observations, actions, and rewards from every device, and distributed actors who take local actions.
    For fair comparison, we use the same actor model on each device and a proportionally scaled critic model for centralized training. 
\end{list}

For the given algorithms, we measure the following metrics for performance evaluation: the number of successfully transmitted packets (Pkt-T), the number of collisions occurred (Pkt-C), the number of lost packets due to buffer saturation (Pkt-L), total network throughput (TPut), the time delay between each successfully transmitted packet (Delay), and normalized gap to measure the fairness across the devices (N-Gap), which is defined as:
\begin{equation}
    \text{N-Gap of }\{x\}=\frac{\max(\{x\})-\min(\{x\})}{\max(\{x\})}
\end{equation}
We consider throughput and packet delay for fairness evaluation.

\begin{table}[!t]
    \captionsetup{skip=6pt}
    \centering
    \setlength\tabcolsep{4pt}
    \caption{Average Network Performance Comparison over Different RA Algorithms. The Proposed Algorithm Achieves Better Performance than the Baselines.}
    \begin{tabular}{c|c|c|c|c|c}
        \hline
        Algorithm & Pkt-T & Pkt-C & Pkt-L & TPut (Mbps) & Delay (ms) \\
        \hline
        RA-P & 4.48 & 5.90 & 5.17 & 39.783 $\pm$ 1.04 & 1.177 $\pm$ 0.06 \\
        RA-ACW & 5.13 & 4.54 & 4.70 & 45.639 $\pm$ 0.81 & 1.235 $\pm$ 0.09 \\
        RA-FCW & 5.63 & 1.54 & 4.09 & 50.006 $\pm$ 0.67 & 0.946 $\pm$ 0.04 \\
        RA-CTDE & \textbf{6.54} & 0.54 & 3.27 & \textbf{58.122} $\pm$ 0.95 & {0.783} $\pm$ 0.01 \\
        Proposed & {6.49} & \textbf{0.50} & \textbf{3.22} & {57.667} $\pm$ 1.39 & \textbf{0.777} $\pm$ 0.02 \\
        \hline
    \end{tabular}
    \label{tab:mean}
\end{table}

\subsection{Results and Discussion}\label{ssec:exp_results}

\subsubsection{Comparison of average performance}

In Table~\ref{tab:mean}, we present a comparison of the average network performance for different RA algorithms.
We observe that the non-RL methods (i.e., RA-P, RA-ACW, and RA-FCW) result in degraded performance across all metrics.
This is primarily due to their probabilistic approach of optimizing performance.
In contrast, both RA-CTDE and our algorithm show significantly improved performance, greatly reducing the number of collisions.
Considering the simulation noise, their performance is nearly identical.
This indicates that our decentralized MARL with local reward sharing is as effective as the CTDE method.
Moreover, our algorithm is far more efficient in terms of overhead complexity and practical applicability.

\begin{figure*}[!t]
    \captionsetup{skip=3pt}
    \centering
    \minipage{0.32\textwidth}
        \includegraphics[width=1\linewidth]{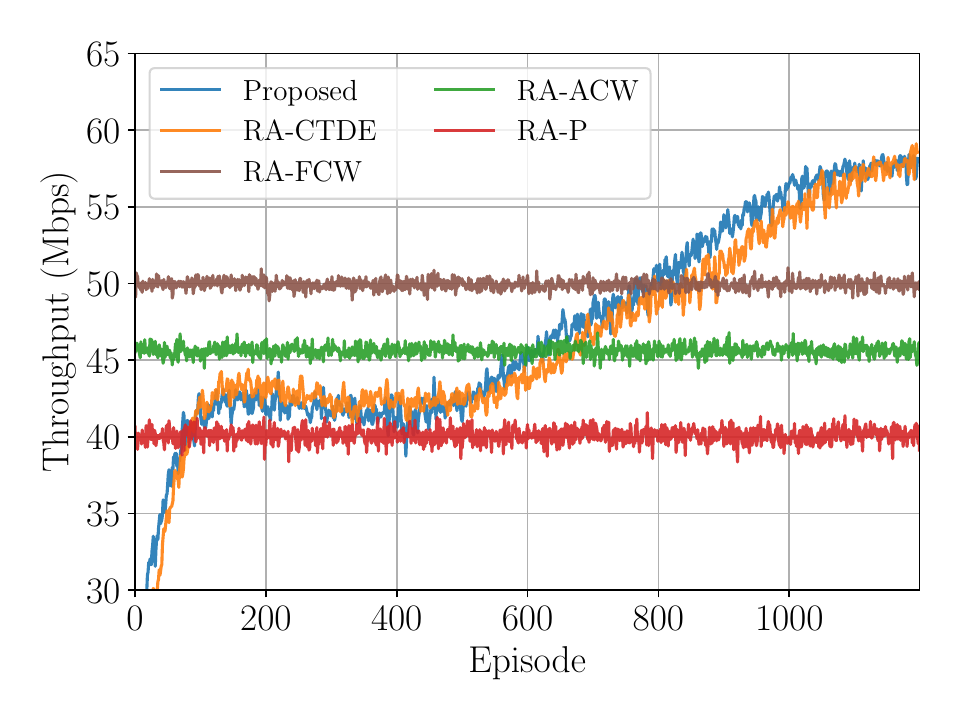}
        \caption{A network throughput versus episode plot for different RA algorithms. While both MARL-based approaches display similar learning pattern, RA-CTDE shows slower convergence.}
        \Description{asdf}
        \label{fig:throughput}
    \endminipage\hfill
    \minipage{0.32\textwidth}
        \includegraphics[width=1\linewidth]{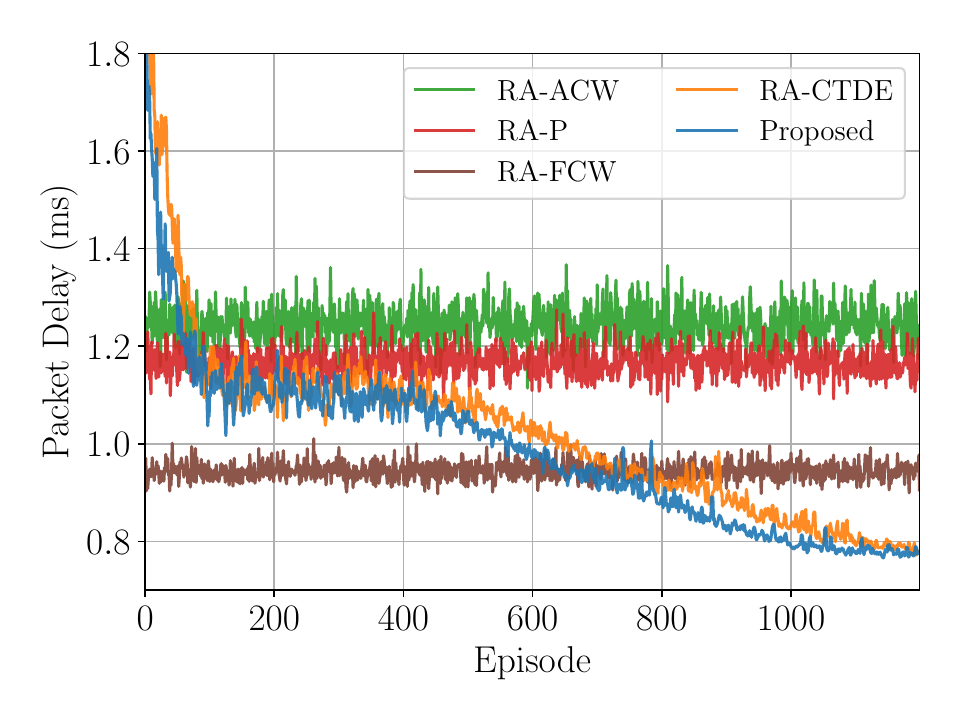}
        \caption{A packet delay versus episode plot for different RA algorithms. RA-ACW yields the worst performance as it avoids collisions by increasing transmission delays.}
        \Description{asdf}
        \label{fig:delay}
    \endminipage\hfill
    \minipage{0.32\textwidth}
        \includegraphics[width=1\linewidth]{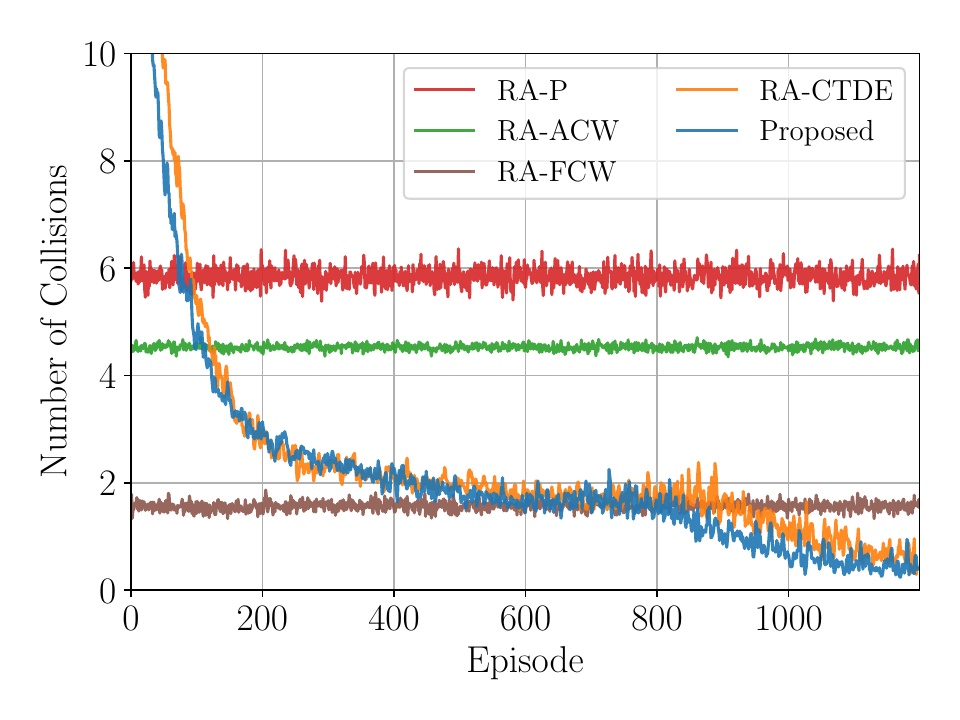}
        \caption{A packet collision frequency versus episode plot for different RA algorithms. Both MARL-based approaches can reduce the rate of collision almost to zero.}
        \Description{asdf}
        \label{fig:collision}
    \endminipage
    \vspace{-1mm}
\end{figure*}

Figure~\ref{fig:throughput} shows the dynamic change in throughput as the learning episode progresses.
Note that non-RL approaches display consistent results throughout the episode as they do not incorporate a process of learning.
We can see that both RA-CTDE and our algorithm improve throughput in a similar manner.
A significant improvement is observed during the first $200$ episodes, primarily due to the reduction in transmission probability to avoid collisions.
After another approximately $800$ episodes of steady performance, the devices converge to a deterministic policy that completely avoids collisions, resulting a steep increase in throughput.
This confirms that the effectiveness of our consensus-based decentralized MARL framework in RA network optimization.
Also, our result verifies that the average consensus on local rewards is sufficient for discovering optimal policies. 
Although the overall learning pattern is similar, it is worth noting that convergence occurs later for RA-CTDE.
This is due to the increased learning complexity of the centralized critic, which processes features of much larger dimensionality.
Figs.~\ref{fig:delay} and~\ref{fig:collision} show the changes in packet delay and collision frequency as learning episodes progress, respectively.
The trends in these figures follow a pattern similar to that in Figure~\ref{fig:throughput}, where RA-CTDE and our algorithm show similar improvements.
Similarly, RA-CTDE converges slower in both metrics.

\subsubsection{Comparison of fairness}

In Table~\ref{tab:min_max}, we provide the maximum and minimum values for throughput and packet delay, along with the normalized gap for each RA algorithm.
Note that a lower gap indicates better fairness.
As shown in the table, non-RL approaches exhibit a large gap in both throughput and packet delay. 
RA-ACW has the largest gap (about a four-time difference in both throughput and delay) likely due to the adaptive contention window introducing unfair delays across devices.
Meanwhile, both RA-CTDE and our algorithm greatly reduce the gap, achieving a much higher fairness level by the end of the learning process.
Note that the improvement is approximately five times compared to the non-RL methods.
The result highlights that, in addition to enhancing throughput performance, our approach can find a policy that maximizes fairness without relying on centralized tasks.

\begin{table}[!t]
    \captionsetup{skip=6pt}
    \centering
    \setlength\tabcolsep{4pt}
    \caption{Network Fairness Comparison over Different RA Algorithms. The Proposed Algorithm Achieves Better Fairness than the Baselines.}
    \begin{tabular}{c|ccc|ccc}
        \hline 
        & \multicolumn{3}{c|}{TPut (Mbps)} & \multicolumn{3}{c}{Delay (ms)} \\
        \hline
        Algorithm & Min & Max & N-Gap & Min & Max & N-Gap \\
        \hline
        RA-P & 5.353 & 14.954 & 0.642 & 0.689 & 1.933 & 0.644 \\
        RA-ACW & 4.588 & 20.116 & 0.772 & 0.555 & 2.339 & 0.772 \\
        RA-FCW & 7.409 & 17.990 & 0.588 & 0.598 & 1.476 & 0.595 \\
        RA-CTDE & 13.128 & 15.256 & \textbf{0.139} & 0.735 & 0.836 & \textbf{0.121} \\
        Proposed & 13.106 & 15.283 & {0.142} & 0.729 & 0.831 & {0.123} \\
        \hline
    \end{tabular}
    \label{tab:min_max}
    \vspace{-0.15in}
\end{table}

\begin{figure}[!t]
    \captionsetup{skip=1pt}
    \centering
    \includegraphics[width=0.92\linewidth]{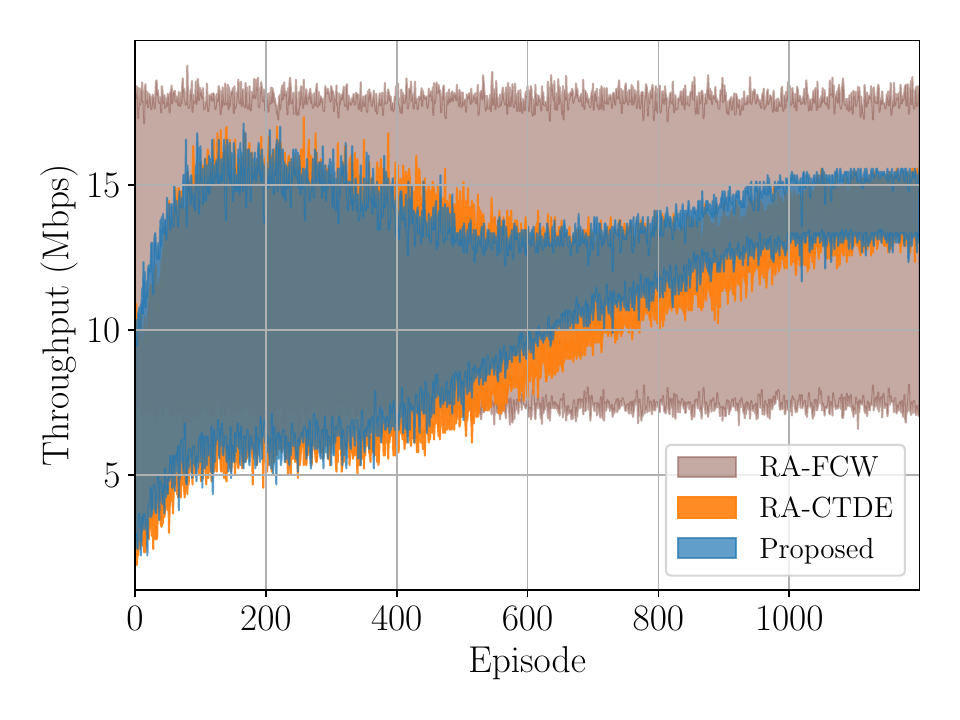}
    \caption{A plot showing the gap between maximum and minimum throughputs. A smaller gap indicates better fairness.}
    \Description{asdf}
    \label{fig:throughput_gap}
    \vspace{-.1in}
\end{figure}

Figure~\ref{fig:throughput_gap} presents a plot showing the change in throughput gap across the learning episodes.
For clarity, we only include RA-FCW from the non-RL approaches.
Although RA-FCW achieves the best fairness among the non-RL algorithms, the throughput gap remains significantly large, with a difference of approximately 10.5 Mbps.
For both MARL-based approaches, the gap is initially similar to that of RA-FCW in the early episodes. 
However, after the $400$th episode, the network begins to learn how to improve fairness.
Starting from the $600$th episode, both the maximum and minimum throughput start to improve together.
Although RA-FCW achieves the highest absolute maximum throughput (around $18$ Mbps), it does so at the expense of sacrificing throughput from other devices.
In contrast, our algorithm learns to prioritize devices with larger packet delays, leading to an overall improvement in total throughput.

\begin{figure}[!ht]
    \captionsetup{skip=1pt}
    \centering
    \includegraphics[width=0.9\linewidth]{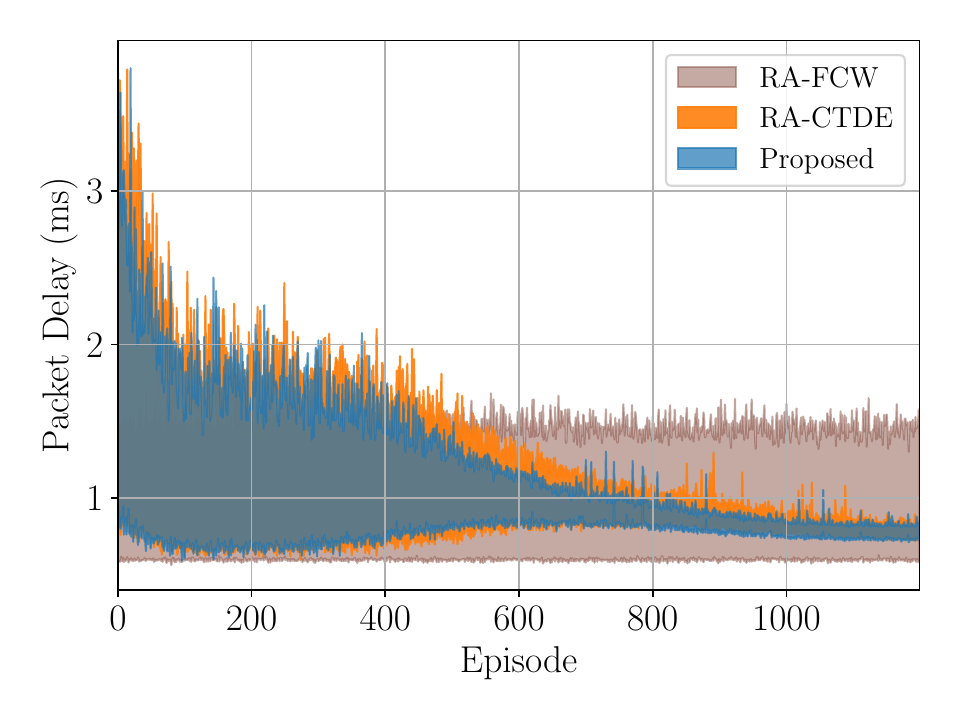}
    \caption{A plot showing the gap between maximum and minimum delays. A smaller gap indicates better fairness.}
    \Description{asdf}
    \label{fig:delay_gap}
    \vspace{-.1in}
\end{figure}

We show the change in delay gap across learning episode for different RA algorithms in Figure~\ref{fig:delay_gap}.
We observe that, for RA-FCW, the gap remains constant throughout across the episodes as no learning is involved.
In the early stages of learning, both RA-CTDE and our algorithm exhibit significant worst-case delays.
By the end of the learning process, both approaches successfully reduce overall packet delays while also improving fairness.
Just like our throughput analysis, it takes longer episodes for RA-CTDE to converge compared to our algorithm.
Despite lacking centralized training, our algorithm is able to successfully discover polices that result in a substantial improvement in RA network performance.

\section{Conclusion}\label{sec:conclusion}

In this paper, we considered the RA-based MAC layer optimization problem and proposed a fully decentralized MARL framework to find RA policies that minimize collisions and ensure transmission fairness across the devices.
After carefully designing MARL parameters, we developed and theoretically analyzed the AC learning performance of our algorithm.
Instead of leveraging centralized training, our algorithm uses the average consensus mechanism to achieve convergence in learning.
Unlike many decentralized MARL strategies, our algorithm only exchanges local rewards, providing a significant advantage in overhead reduction.
We showed through theoretical and numerical analysis that our proposed approach can attain network performance comparable to that of CTDE. 

\bibliographystyle{ACM-Reference-Format}
\bibliography{mybib}

\end{document}